\documentclass{jetpl}
\twocolumn
\title{Topological invariant  for superfluid  $^3$He-B and quantum phase transitions}

\rtitle{Topological invariant  for superfluid  $^3$He-B and quantum phase transitions}

\sodtitle{Topological invariant  for superfluid  $^3$He-B and quantum phase transitions}

\author{
G.E. Volovik
 \/\thanks{
volovik@boojum.hut.fi} 
}

\rauthor{
G.E.Volovik
}

\sodauthor{
Volovik
}

\address{Low Temperature Laboratory, Helsinki University of
Technology, P.O.Box 5100, FIN-02015, HUT, Finland
\\
 Landau Institute for Theoretical Physics RAS, Kosygina 2,
119334 Moscow, Russia}

\dates{September 22, 2009}{*}

\abstract{
We consider topological invariant describing the vacuum states of superfluid $^3$He-B, which belongs
to the special class of  time-reversal invariant topological insulators and superfluids.  Discrete symmetries important for classification of the topologically distinct vacuum states are discussed. 
One of them leads to the additional subclasses of $^3$He-B states and is responsible for the finite density of states of Majorana fermions  living at some interfaces between the bulk states.   Integer valued topological invariant is expressed in terms of the Green's function, which allows us to consider systems with interaction. 
}

\begin{document}

\maketitle


\section{Introduction}

General classification schemes based on topology \cite{Schnyder2008,Schnyder2009a,Schnyder2009b,Kitaev2009,Volovik2003,Volovik2007,Horava2005} suggest existence of the topological insulators and fully gapped topological superfluids/superconductors which have the gapless edge states on the boundary or at the interface. 
Superfluid $^3$He-B  belongs to the special class of  three-dimensional  topological superfluids with time-reversal symmetry. Topological invariant which describes the ground states (vacua) of  $^3$He-B has been discussed in Refs. \cite{Schnyder2008} and \cite{SalomaaVolovik1988}.
Here we present the explicit expression for the relevant topological invariant, and discuss topological quantum phase transitions occurring between the vacuum states and fermion zero modes living at the interfaces between the bulk states.

\section{Topological invariant protected by symmetry}

Usually in the topological classification of ground states people use the Hamiltonian
of free particles or the corresponding effective Hamiltonian such as Dirac and Bogoliubov-de Gennes Hamiltonians \cite{Schnyder2008,Schnyder2009a,Schnyder2009b,Kitaev2009}.  
However, in this classification the natural problem arises, what is the effect of interaction between particles. Moreover, the original first-principle many-body Hamiltonian of, say, liquid $^3$He
\begin{eqnarray}
{\cal H}-\mu {\cal N}=\int d{\bf x}\psi^\dagger({\bf x})\left[-{\nabla^2\over 2m}
-\mu
\right]\psi({\bf x}) 
\nonumber
\\
+{1\over 2}\int d{\bf x}d{\bf y}U({\bf x}-{\bf
y})\psi^\dagger({\bf x})
\psi^\dagger({\bf y})\psi({\bf y})\psi({\bf x}),
\label{TheoryOfEverything}
\end{eqnarray}
has no information on the topological structure of the ground state of the system --  superfluid $^3$He-B.
The accurate procedure to reduce such a strongly interacting many-body system to the effective coarse-grained Hamiltonian is absent. However, the microscopic Hamiltonian allows us at list in principle to calculate the Green's function $G(\omega,{\bf p})$ -- the quantity which determines the main properties of the  translational invariant or periodic ground states of the system.
That is why the object for the topological classification must be the Green's function rather than Hamiltonian. Then it is applicable even in cases when one cannot introduce the effective low energy Hamiltonian, for example when  Green's function does not have poles, see \cite{Volovik2007,FaridTsvelik2009}. 

Green's function topology has been used in particular  for classification of topologically protected nodes in the quasiparticle energy spectrum of systems of different dimensions; for the classification of the topological ground states in the fully gapped $2+1$ systems, which experience intrinsic quantum Hall and spin-Hall effects  \cite{VolovikYakovenko1989,Yakovenko1989,SenguptaYakovenko2000,Volovik2003,Volovik2007,ReadGreen2000}, including multi-band topological insulators \cite{Volovik1988};   in relativistic quantum field theory  of $2+1$ massive Dirac fermions 
  \cite{IshikawaMatsuyama1986,IshikawaMatsuyama1987,Jansen1996}; etc.

The integer-valued topological invariants were expressed via the Green's function, which was considered at imaginary frequency to avoid the zeroes, poles or other possible singularities on the mass shell.
In the fully gapped  systems one may consider  the Green's function not only at imaginary frequency, but also at  real frequency if it is below the gap. For the classification of the $^3$He-B states we shall use the Green's function at zero frequency:
 ${\cal G}({\bf p})\equiv G(\omega=0,{\bf p})$. 
The typical example of the integer valued  topological invariant is the following 3-form:
 \begin{equation}
N = {e_{ijk}\over{24\pi^2}}~
{\bf tr}\left[\int   d^3p 
~ {\cal G}\partial_{p_i} {\cal G}^{-1}
{\cal G}\partial_{p_j} {\cal G}^{-1} {\cal G}\partial_{p_k}  {\cal
G}^{-1}\right]
\,,
\label{3DTopInvariantGen}
\end{equation}
where integration is over the whole momentum space for translational invariant systems, or over the Brillouin zone in crystals.
For $^3$He-B obeying  time-reversal symmetry this invariant is identically zero,
$N=0$. However, the discrete symmetries of  $^3$He-B give rise to the other invariants.
Examples of additional  integer valued  topological invariants which appear due to symmetry  in different condensed matter systems and in quantum field theory can be found in Refs. \cite{VolovikYakovenko1989,Volovik1992,SenguptaYakovenko2000,Volovik2003,Volovik2007}.

Due to  symmetry,  one or several Pauli matrices $\gamma$ of spin, pseudospin or Bogoliubov-Nambu spin may either commute or anti-commute with the Green's function matrix:
\begin{equation}
\gamma {\cal G}({\bf p})=-{\cal G}({\bf p})\gamma \,,
\label{eq:symmetry1}
\end{equation}
or
\begin{equation}
\gamma {\cal G}({\bf p}) ={\cal G}({\bf p})\gamma
\,.
\label{eq:symmetry2}
\end{equation}
This leads to the following integer valued  topological invariants:
\begin{equation}
N_\gamma = {e_{ijk}\over{24\pi^2}} ~
{\bf tr}\left[\gamma \int   d^3p 
~ {\cal G}\partial_{p_i} {\cal G}^{-1}
{\cal G}\partial_{p_j} {\cal G}^{-1} {\cal G}\partial_{p_k}  {\cal
G}^{-1}\right]\,,
\label{3DTopInvariant_mu}
\end{equation}
with ${\cal G}({\bf p})$ and $\gamma$ obeying either \eqref{eq:symmetry1}
 or \eqref{eq:symmetry2}.

The topological classes of the $^3$He-B states can be represented by the following Green's function, which plays the role of  effective Hamiltonian:
 \begin{eqnarray}
{\cal G}^{-1}({\bf p})=M(p)\tau_3+  \tau_1\left( \sigma_x c_xp_x + \sigma_y c_yp_y + \sigma_z c_zp_z   \right)
\,,
\label{eq:B-phase}
\\
M(p)=\frac{p^2}{2m^*} - \mu~,
\label{eq:B-phase2}
\end{eqnarray}
where $\tau_i$ are Pauli matrices of Bogolyubov-Nambu spin. 
The overall `conformal' factor, which may depend on ${\bf p}$,  is omitted since it does not influence
the topological invariant.
In the isotropic $^3$He-B all `speeds of light' are equal, $|c_x|= |c_y|=|c_z|=c$. The topological invariant relevant  for $^3$He-B is $N_\gamma$ in \eqref{3DTopInvariant_mu} with $\gamma=\tau_2$:
\begin{equation}
N_\gamma = {e_{ijk}\over{24\pi^2}} ~
{\bf tr}\left[\tau_2 \int   d^3p 
~ {\cal G}\partial_{p_i} {\cal G}^{-1}
{\cal G}\partial_{p_j} {\cal G}^{-1} {\cal G}\partial_{p_k}  {\cal
G}^{-1}\right]\,.
\label{3DTopInvariant_tau}
\end{equation} 
 The $\tau_2$ matrix plays the role of the 
$\gamma$-matrix, which anti-commutes with the Green's function in \eqref{eq:symmetry1}. The $\tau_2$ symmetry   is combination of time reversal  and particle-hole symmetries in $^3$He-B.

 \section{Phase diagram in mass plane}
 
 \begin{figure}[top]
\centerline{\includegraphics[width=1.0\linewidth]{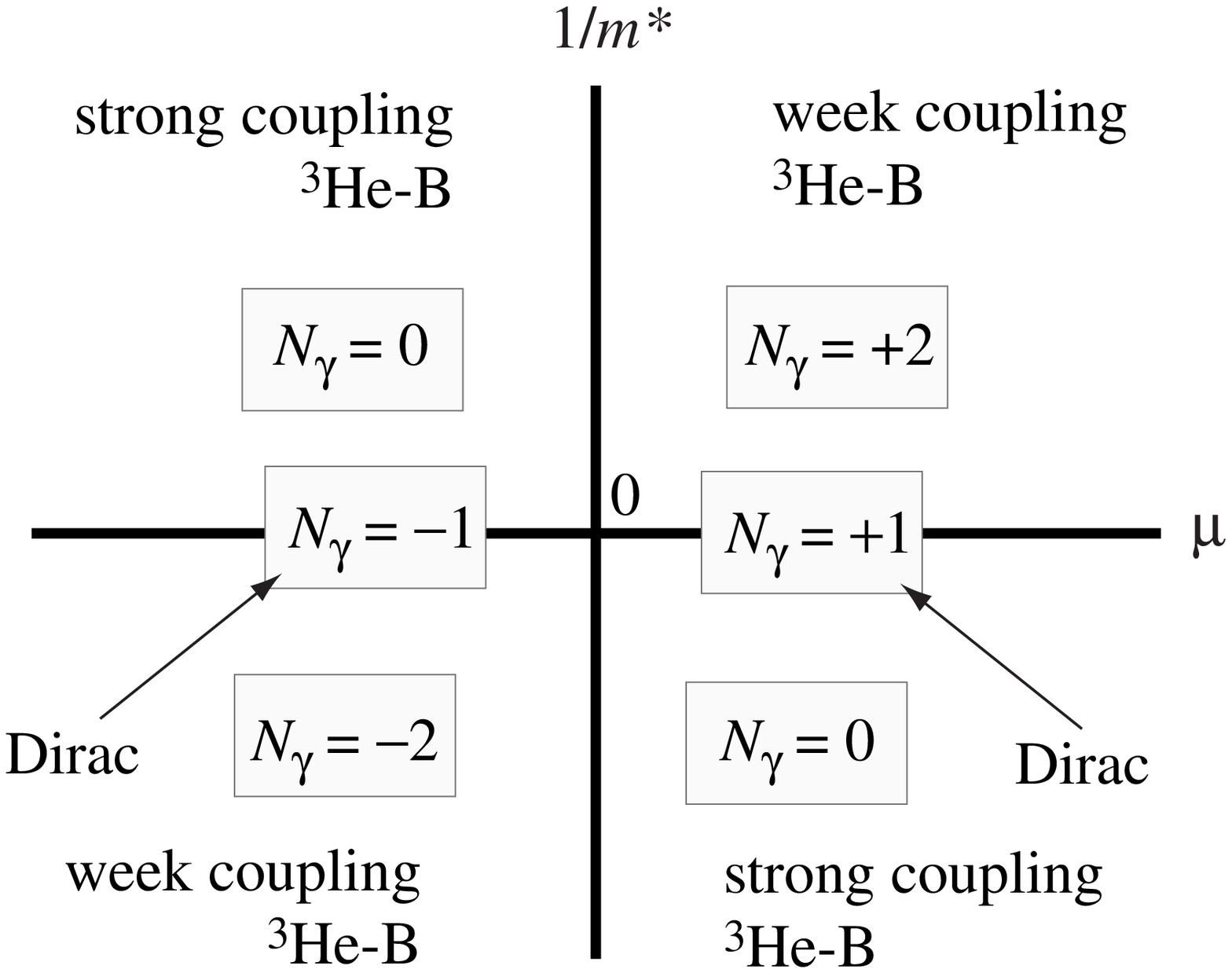}}
  \caption{\label{QPT}  Phase diagram of topological states of $^3$He-B in equation \eqref{eq:B-phase} in the plane $(\mu,1/m^*)$ for the speeds of light $c_x>0$, $c_y>0$ and $c_z>0$. States on the line 
  $1/m^*=0$ correspond to the  Dirac vacua, which Hamiltonian is non-compact. Topological charge of the Dirac fermions  is intermediate between charges of compact $^3$He-B states.
The line $1/m^*=0$ separates the states with different asymptotic behavior of the Hamiltonian at infinity:
${\cal G}^{-1}({\bf p}) \rightarrow \pm \tau_3 p^2/2m^*$. 
 The line $\mu=0$ marks topological quantum phase transition, which occurs between the weak coupling $^3$He-B  (with $\mu>0$, $m^*>0$ and topological charge $N_\gamma=2$) and the strong coupling $^3$He-B   (with $\mu<0$, $m^*>0$ and $N_\gamma=0$).    This transition is topologically equivalent to quantum phase transition between Dirac vacua with opposite mass parameter 
 $M=\pm |\mu|$, which occurs when $\mu$ crosses zero along the line $1/m^*=0$. 
 The interface which separates two states contains single Majorana fermion in case of $^3$He-B, and single chiral fermion in case of  relativistic quantum fields.  Difference in the nature of the fermions is that in Bogoliubov-de Gennes system the components of spinor are related by complex conjugation. This reduces the number of degrees of freedom compared to Dirac case.
 }
\end{figure}

Fig. \ref{QPT} shows the phase diagram of topological states of $^3$He-B in the plane $(\mu,1/m^*)$.
On the line  $1/m^*=0$ one obtains Dirac fermions with mass parameter $M=-\mu$:
 \begin{equation}
{\cal G}^{-1}({\bf p})=M\tau_3+  \tau_1\left( \sigma_x c_xp_x + \sigma_y c_yp_y + \sigma_z c_zp_z   \right)
\,.
\label{eq:Dirac}
\end{equation}
 Topological invariant \eqref{3DTopInvariant_tau} for Dirac fermions is
  \begin{equation}
N_\gamma=-{\rm sign}(Mc_x c_y c_z)={\rm sign}(\mu c_x c_y c_z)
\,.
\label{eq:DiracInvariants}
\end{equation}
In relativistic quantum field theory, $\tau_2$ matrix in \eqref{3DTopInvariant_tau} is the operator of $CT$ symmetry \cite{WeinbergBook}.
Hamiltonian for Dirac fermions is non-compact, with different asymptotes at $p\rightarrow \infty $ for different directions of momentum ${\bf p}$. As a result, the topological charge of the Dirac fermions  is intermediate between charges of the compact states of $^3$He-B below and above the horizontal axis
(see Refs. \cite{Haldane1988}, \cite{Schnyder2008} and \cite{Volovik2003} on the marginal behavior of fermions with relativistic spectrum; also note that   the topological invariant $N_\gamma$ in
\eqref{3DTopInvariant_tau} has values twice larger than the invariants introduced in  Refs. 
\cite{Schnyder2009b} and  \cite{SalomaaVolovik1988}).
When the line  $1/m^*=0$ is crossed, the asymptotic behavior of the $^3$He-B Green's function changes from ${\cal G}^{-1}({\bf p}) \rightarrow  + \tau_3 p^2/2m^*$ to ${\cal G}^{-1}({\bf p}) \rightarrow  -\tau_3 p^2/2m^*$.

 The real superfluid $^3$He-B lives on the weak coupling side of the phase diagram: at $\mu>0$, $m^*>0$, $\mu\gg m^*c^2$. However, in the ultracold Fermi gases with triplet pairing
 the strong coupling limit is possible near the Feshbach resonance \cite{GurarieRadzihovsky2007}.
 When $\mu$ crosses zero the topological quantum phase transition occurs, at which the topological charge $N_\gamma$ changes from  $N_\gamma=2$ to  $N_\gamma=0$. 
 Previously the topological quantum phase transition occurring at $\mu=0$ has been considered
for the chiral $p$-wave states of the $^3$He-A type: the transition from the gapless state at $\mu>0$ to the fully gapped state at $\mu<0$ \cite{Volovik1992,GurarieRadzihovsky2007}. In the $^3$He-B case, both states are fully gapped, while the intermediate state at $\mu=0$ is gapless: it has two point nodes at 
 ${\bf p}=0$ with opposite chiralities.  There is a general relation between topological invariants of the two states and the number of point nodes in the intermediate gapless state  \cite{Volovik2003}, which for a given case reads:
\begin{equation}
N_{\rm point~nodes} = |N_\gamma(\mu >0)-N_\gamma(\mu<0)|=2 
\,.
\label{connection}
\end{equation}
This relation also determines the number of $2+1$ fermion zero modes living at the interface between the two states:
\begin{equation}
N_{\rm FZM}=\frac{1}{2}\left|N_\gamma(\mu >0)-N_\gamma(\mu<0)\right| =1  
\,.
\label{FZMnumber}
\end{equation}
This analog of the index theorem \cite{JackiwRossi1981}  implies that such interface contains single Majorana fermion.

The same quantum phase transition occurs when $\mu$ crosses zero along the line $1/m*=0$, i.e. the line of the relativistic Dirac fermions. At this transition the mass $M=-\mu$ of Dirac fermions changes sign.
The rules \eqref{connection} and \eqref{FZMnumber} are also applicable for the Dirac fermions.
However, in relativistic quantum field theory the domain wall separating vacua with opposite $M$ contains chiral fermion rather than the Majorana one.  Difference in the nature of the fermions comes from the observation that in the Bogoliubov-de Gennes system the components of a spinor are related by complex conjugation. This reduces the number of fermionic degrees of freedom compared to the  relativistic quantum field theory and makes them the Majorana fermions.

The spectrum of fermion zero modes at this interface can be easily found in the limit $|\mu| \ll m^*c^2$, when the $p^2$ term in the Hamiltonian can be neglected and the Hamiltonian approaches the Dirac one. In this limit Majorana fermions in $^3$He-B have relativistic spectrum:
 \begin{equation}
H_{\rm zm} = c \hat{\bf z} \cdot({\mbox{\boldmath$\sigma$}} \times {\bf p})~.
\label{eq:ModesH}
\end{equation} 
The same spectrum is obtained for the chiral fermions living on the interface between the states of the
3D topological insulator with normal and inverted  arrangements of two bands, see Ref. \cite{VolkovPankratov1985}.

 The invariant \eqref{3DTopInvariant_tau} is applicable to the general case of systems with Green's function of the type
 \begin{equation}
{\cal G}^{-1}({\bf p})=\hat\epsilon({\bf p}) \tau_3+\hat\Delta({\bf p})  \tau_1
\,,
\label{eq:Hamiltonian}
\end{equation}
where $\hat\epsilon$ and $\hat\Delta$ are Hermitian matrices with spin and band indices. 
This includes in particular the Hamiltonians discussed in Refs. \cite{Schnyder2009b} and \cite{Sato2009}. The phase  diagram of topological 3D states in \cite{Schnyder2009b} is similar to that in Fig. \ref{QPT} (note again that the topological invariant 
$N_\gamma$ has values twice larger than invariants introduced in  Refs. 
\cite{Schnyder2009b} and  \cite{SalomaaVolovik1988}).

\section{Phase diagram in the  speed  of light plane}

\begin{figure}[top]
\centerline{\includegraphics[width=1.0\linewidth]{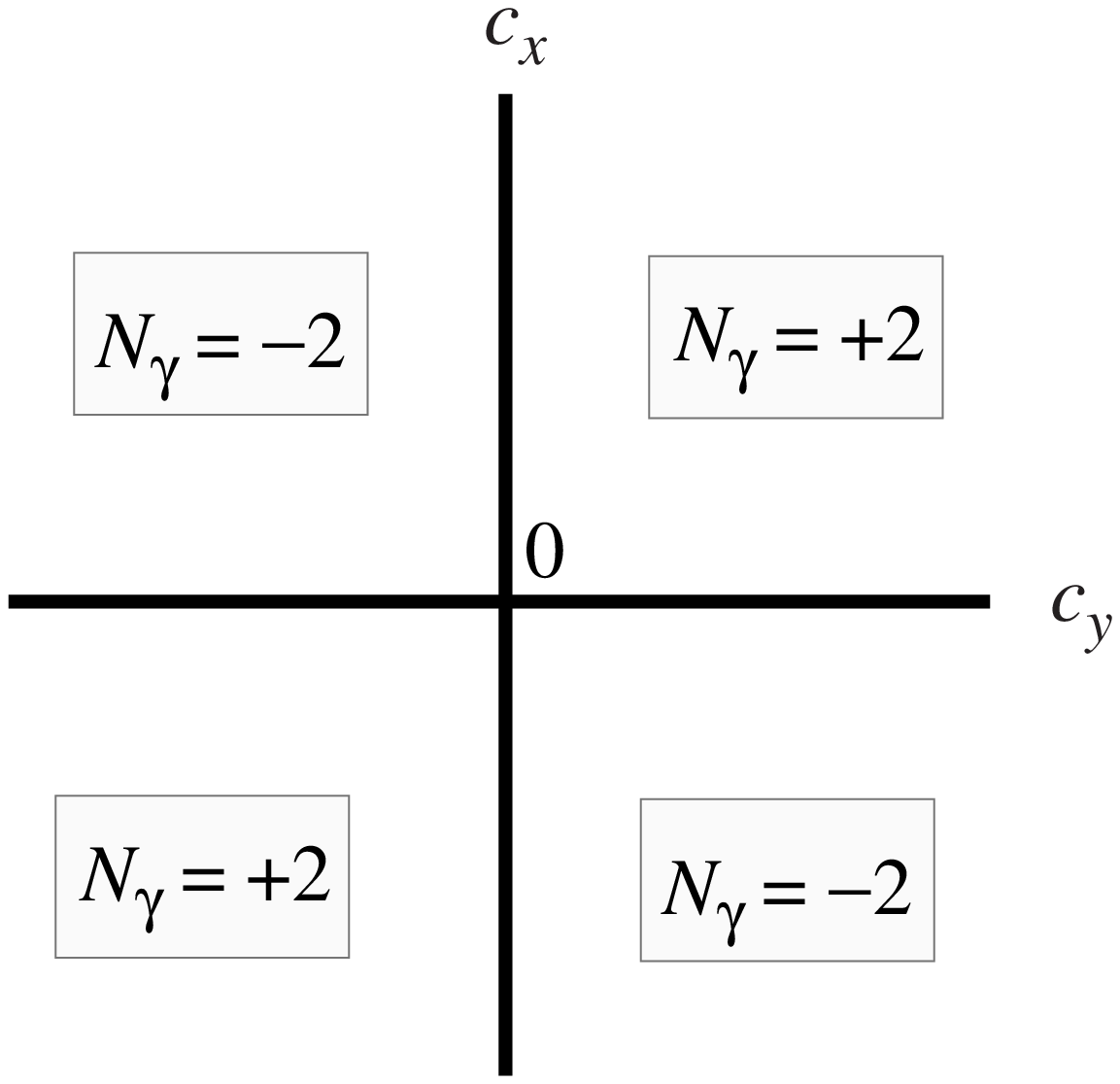}}
  \caption{\label{PD}  Phase diagram of $^3$He-B states at fixed $c_z>0$, $\mu>0$ and $m^*>0$.
  The interface between the states with different winding number $N_\gamma$ contains Majorana fermions. However,  in the presence of  discrete symmetry $Z_2$ -- combined rotation by $\pi$ about $z$ axis
  in spin and orbital spaces --  the $^3$He-B states with the same topological invariant $N_\gamma$
 may be connected only via the gapless state. As a result, the interface between two states with the same $N_\gamma$ also contains Majorana fermions. These fermions have finite density of states at zero energy \cite{SalomaaVolovik1988}. This is the origin of the finite density of states
of the Majorana edge states on the diffusive wall \cite{Volovik2009}. 
}
\end{figure}

Fig. \ref{PD} shows phase diagram of topological states of $^3$He-B in the plane $(c_x,c_y)$
 at fixed $c_z>0$, $\mu>0$ and $m^*>0$.
When one of the components of speed of light, say,  $c_x$ crosses zero, two pairs of point nodes appear in the intermediate gapless state, at points ${\bf p}=\pm(0,0,p_F)$, where $p_F^2=2\mu m^*$
is Fermi momentum. This is the consequence of different topological invariant $N_\gamma$ of the two states:
\begin{equation}
N_{\rm point~nodes} = |N_\gamma(c_x >0)-N_\gamma(c_x<0)|=4 
\,.
\label{connection2}
\end{equation}
  The intermediate state is the so called planar state \cite{VollhardtWolfle}, and  the corresponding gapless fermions in this state are characterized by the topological invariant protected by discrete symmetry in \eqref{eq:symmetry2} $\gamma=\tau_3\sigma_x$, which commutes with Hamiltonian: $\gamma {\cal G}({\bf p})={\cal G}({\bf p})\gamma$ (see Eq.(14.4) of Ref. \cite{Volovik2003}):
 \begin{equation}
N_3 = {e_{\mu\nu\lambda\gamma}\over{24\pi^2}}
{\bf tr}\left[\tau_3\sigma_x\int   dS^\gamma 
 G\partial_{p_\mu} G^{-1}
G\partial_{p_\nu} G^{-1} G\partial_{p_\lambda} G^{-1}\right],
\label{FPTopInvariantP}
\end{equation}
where integral is around a nodal point $p_{\mu}=(0,0,p_F,0)$ or $p_{\mu}=(0,0,-p_F,0)$ in the 4D momentum-frequency space $p_{\mu}=({\bf p},\omega)$.
 Number of Majorana fermion zero modes  living at the interface between the two states is 
 correspondingly twice smaller: $N_{\rm FZM}=2$. The same number of $2+1$ fermions live at the boundary of $^3$He-B with specular reflection. This is because there is an exact mapping between the edge states on the $^3$He-B boundary with perfect reflection and fermion zero modes living at the interface, see Ref.  \cite{Volovik2009}.
  
There are subclasses of $^3$He-B states, which exist due to discrete symmetry.
  When two components of speed of light, say,  $c_x$ and $c_y$ cross  zero, the 
 topological charge $N_\gamma$ does not change. The final state can be continuously connected with original state by spin rotation by angle $\pi$ about axis $z$.
However, there is additional discrete symmetry  -- the combined rotation by $\pi$ about $z$ axis
  in spin and orbital spaces. If this $Z_2$  symmetry is maintained, the two states with the same topological invariant $N_\gamma$  cannot be connected by adiabatic deformations. The intermediate gapless states necessarily contains the line of nodes. 
 
As a result the interface between states with the same $N_\gamma$, at which  two components of speed of light
in \eqref{eq:DiracInvariants} change sign, also contains zero modes. Spectrum of these fermions $H_{\rm zm} \propto  c \sigma_xk_y$, and they  have finite density of states at zero energy \cite{SalomaaVolovik1988,Volovik2009}. Because of the mapping between  the fermion zero modes living at this interface and the edge state of $^3$He-B on diffusive boundary, the Majorana states on the diffusive wall also have finite density of states \cite{Volovik2009} in agreement with the detailed quasiclassical theory \cite{Kopnin1991}. 

\section{Discussion}

There are many experimental evidences for  the Andreev surface states on the wall of $^3$He-B, see e.g.  \cite{CCGMP,Davis2008,Nagai2008,Murakawa2009a}, and some other experiments are proposed \cite{ChungZhang2009}. At low temperature, where thermal quasparticles in bulk $^3$He-B are frozen out exponentially, the low-energy surface states -- Majorana fermions --  will give the main contribution to thermodynamics and dissipation, with the power-law dependence of the physical quantities. The extra discrete symmetry of $^3$He-B essentially enhances the effect of Andreev-Majorana states by providing the finite density of states of these modes on rough walls \cite{Murakawa2009b}. Systems with highly developed surfaces, such as $^3$He-B confined in porous materials, or systems of parallel plates or tubes, would be more advantageous for the direct identification of the Majorana fermions in superfluid $^3$He-B. 
 
It is a pleasure to thank  A. Kitaev and Yu. Makhlin  for valuable discussions. This work is supported in part  by the Academy of Finland,
Centers of Excellence Program 2006--2011,
the Russian Foundation for Basic Research (grant 06--02--16002--a),
and the Khalatnikov--Starobinsky leading scientific school (grant 4899.2008.2).


\end{document}